\documentclass[aps,preprint]{revtex4}%
\usepackage{amsfonts}
\usepackage{amsmath}
\usepackage{amssymb}
\usepackage{graphicx}%
\providecommand{\U}[1]{\protect\rule{.1in}{.1in}}

\begin{document}
\title{ Kinetic Theory for Matter Under Extreme Conditions}
\author{James Dufty and Jeffrey Wrighton}
\affiliation{Department of Physics, University of Florida}

\begin{abstract}
The calculation of dynamical properties for matter under extreme conditions is
a challenging task. The popular Kubo-Greenwood model exploits elements from
equilibrium density functional theory (DFT) that allow a detailed treatment of
electron correlations, but its origin is largely phenomenological; traditional
kinetic theories have a more secure foundation but are limited to weak
ion-electron interactions. The objective here is to show how a combination of
the two evolves naturally from an exact short time limit for the generator of
the effective single electron dynamics governing time correlation functions.
This provides a theoretical context for the current DFT-related approach, the
Kubo-Greenwood model, while showing the nature of its corrections. The method
is to calculate the exact short time dynamics in the single electron subspace,
for a given configuration of the ions. This differs from the usual kinetic
theory approach in which an average over the ions is performed as well. In
this way the effective ion-electron interaction is treated exactly and shown
to be determined from DFT. The correlation functions have the random phase
approximation for an inhomogeneous system, but with renormalized ion-electron
and electron-electron potentials. The dynamic structure function, density
response function, and electrical conductivity are calculated as examples. The
static local field corrections in the dielectric function are identified in
this way. The current analysis is limited to semi-classical electrons (quantum
statistical potentials), so important quantum conditions are excluded.
However, a quantization of the kinetic theory is identified for broader
application while awaiting its detailed derivation.

\end{abstract}
\date{\today}
\maketitle

\section{Introduction}

\label{sec1}Recent interest in matter under extreme conditions (e.g., high
pressure materials, warm dense matter, and high temperature plasmas) has been
stimulated in part by new experimental access to such states \cite{Glenzer16}.
Thermodynamic properties can be addressed in a controlled way by finite
temperature density functional theory (DFT) for the electrons \cite{DFTrev} in
conjunction with \textit{ab initio }molecular dynamics (AIMD) for the ions
\cite{AIMD}. The two tools of DFT and MD are appropriate for such states since
strong coupling and bound states can be treated without explicit
approximation. Dynamical phenomena are more problematical, particularly for
the electrons which require quantum dynamics in general. Current many-body
theories of condensed matter physics or plasma physics have questionable
applicability. Kinetic theories, classical or quantum, typically have the
limitation to free electrons and weak ion-electron correlations. Models which
utilize DFT quantities (such as the Kubo-Greenwood model \cite{KG} below)
remove this restriction at the price of phenomenology and unknown context. The
objective here is to demonstrate a kinetic theory for electrons in a given
configuration of ions, obtained in an exact limit that incorporates the DFT
model and well-known many-body effects of electron dynamics. The derivation is
semi-classical but makes no explicit restrictions regarding coupling among the
ions and electrons, or bound and free electron states.

The system of interest here is that of electrons in equilibrium with a frozen
disordered configuration of ions in the grand canonical ensemble. In
applications, properties for this system are found by averaging over different
ionic configurations. The properties of interest are time correlation
functions that determine transport coefficients, scattering cross sections,
and other equilibrium dynamic electron fluctuations of linear response. The
state conditions include possible strong coupling and other effects of
correlations such as dynamical fluctuations. These are the types of difficult
cases that are handled well by DFT for thermodynamic calculations, and it is
tempting to think (hope) that such detailed information can be exploited
somehow for the analysis of dynamics as well. A practical implementation of
this idea, the Kubo-Greenwood model (KG) \cite{KG}, is obtained by replacing
the many-body electron-ion Hamiltonian by a sum of single particle
Hamiltonians. Though in principle arbitrary, invariably they are chosen to be
non-interacting particles governed by the Kohn-Sham potential which includes
the external ion field. The latter appears in the variational Euler equation
of DFT and is determined from the exchange-correlation free energy as a
functional of the equilibrium density \cite{DFTrev}. In the KG model the
Kohn-Sham Hamiltonian, originally defined to determine the equilibrium
density, is promoted to a generator for the dynamics. Furthermore, it is
assumed to represent in a mean field way the effects of the true Coulomb
interactions which it replaces. The origin of this picture of Kohn-Sham
quasi-particles and the conditions for its validity have not been established.
Doing so is a primary objective of the present work.

In earlier work the authors considered time correlation functions for a
semi-classical electron gas in the presence of a single fixed ion, and derived
a kinetic equation whose form is exact at short times \cite{Wrighton08}. It
has a single particle dynamics with a "renormalized" external potential from
the electron - ion interaction. In addition, there is a collective Vlasov
(random phase approximation) dynamics with a renormalized electron - electron interaction. Both of
these effective interactions are determined from derivatives of the
semi-classical exchange - correlation free energy. In particular, the electron
- ion interaction is precisely the Kohn-Sham potential of the KG model. Here,
that description is extended to time correlation functions for electrons in a
disordered many-ion background with configurations sampled from an equilibrium
ensemble. An important distinguishing feature of this kinetic equation
relative to others is that only electron degrees of freedom are averaged out
to define the single electron subspace. No ion average is performed. This
allows a detailed description of the average electron interaction with each
ion, setting the framework for connection with DFT. A brief summary of these
ideas is given in reference \cite{Dufty17}.

The exact analysis of the generator for the dynamics at short times at present
has been established only for semi-classical electrons. Thus strong degeneracy
and other extreme quantum effects are not captured in the exact analysis here.
This limits the conditions for applicability, as described briefly in section
\ref{sec6}. However, the primary objective here is to show by example how the
many-body problem can be controlled, and to indicate a pathway to the basis
for the KG model. In section \ref{sec5} a straightforward quantization of this
semi-classical result is given as a temporary "placeholder" for the detailed
quantum derivation paralleling that given here.

In the next section, the time correlation functions and their representation
in terms of linear kinetic theory are introduced. The formal definition of the
generator for the electron dynamics in the single electron subspace is
defined. While generally time dependent, its exact form is evaluated at $t=0$
in Appendix \ref{appA}. No limitations are placed on the strength of
correlations or coupling. This gives a Markov dynamics for the time
correlation functions in which the generator is taken to have the same form at
later times as well. The result has the structure of a kinetic theory in the
random phase approximation, extended to an inhomogeneous system due to the
external forces of the ions. However, that force is renormalized to be derived
from the Kohn-Sham potential (the ion field plus the first functional
derivative of the excess free energy), and the electron-electron Coulomb
potential is renormalized to the electron-electron direct correlation function
(second functional derivative of the excess free energy). As illustrations,
the dynamical structure function, density response function, and frequency
dependent electrical conductivity are determined from this kinetic equation.
The dielectric function is considered, and the associated static local field
corrections are identified. In the last section these results are summarized
and discussed.

\section{Time Correlation Functions from Kinetic Theory}

\label{sec2}Consider a system of $N_{e}$ electrons in equilibrium with $N_{i}$
fixed ions. The Hamiltonian is%

\begin{equation}
H=\sum_{\alpha=1}^{N_{e}}\left(  \frac{1}{2}mv_{\alpha}^{2}+V_{ei}%
(\mathbf{r}_{\alpha},\left\{  \mathbf{R}\right\}  )\right)  +\frac{1}{2}%
\sum_{\alpha\neq\gamma=1}^{N_{e}}V_{ee}(r_{\alpha\gamma}) \label{2.1}%
\end{equation}
where the interaction potential for the electrons with the $N_{i}$ fixed ions
is%
\begin{equation}
V_{ei}(\mathbf{r}_{\alpha},\left\{  \mathbf{R}\right\}  )\equiv\sum_{\gamma
=1}^{N_{i}}V_{ei}(\left\vert \mathbf{r}_{\alpha}-\mathbf{R}_{\gamma
}\right\vert ). \label{2.2}%
\end{equation}
The notation $\left\{  \mathbf{R}\right\}  $ denotes a dependence on the
collection of $N_{i}$ fixed ion coordinates $\mathbf{R}_{\gamma}$. Also,
$\mathbf{r}_{\alpha}$ and $\mathbf{v}_{\alpha}$ are the position and velocity
of electron $\alpha$. In the quantum case all interactions are pure Coulomb.
The analysis below is entirely within classical mechanics for both ions and
electrons. Residual quantum effects must be retained to prevent collapse due
to the electron-ion singularity at zero separation. In the quantum case such
collapse is avoided due to diffraction effects. These can be accounted for in
the classical representation by regularizing the Coulomb potential within a
distance of the order of the thermal de Broglie wavelength. Similar effects
occur for the electron-electron interaction, which also has an additional
effect due to Pauli exclusion. The use of such modified potentials has a long
history, leading to many different forms originating from different contexts
for their derivation \cite{Jones07}. The resulting classical representation
allows MD simulation of opposite charge components. One of the first was an
application to hydrogen plasmas \cite{Hansen80}; a more recent simulation in
the current context is that of references \cite{Whitley15,Benedict12}. The
specific forms, or their limitations, are not central to the discussion here.
Instead, the objective is to demonstrate how the many-body physics can be
analysed in a controlled way to make contact with current phenomenology and to
clarify its context.

The equilibrium time correlation functions for two observables $A$ and $B$ in
the grand canonical ensemble are
\begin{equation}
\left\langle A(t)\delta B;\left\{  \mathbf{R}\right\}  \right\rangle
=\sum_{N_{e}}\int d\left\{  x\right\}  \rho_{e}\left(  \left\{  \mathbf{R}%
\right\}  ,\left\{  x\right\}  \right)  A(t,\left\{  \mathbf{R}\right\}
,\left\{  x\right\}  )\delta B\left(  \left\{  \mathbf{R}\right\}  ,\left\{
x\right\}  \right)  , \label{2.9}%
\end{equation}%
\begin{equation}
\rho_{e}\left(  \left\{  \mathbf{R}\right\}  ,\left\{  x\right\}  \right)
=e^{\beta\Omega\left(  \left\{  \mathbf{R}\right\}  \right)  }e^{-\beta\left(
H\left(  \left\{  \mathbf{R}\right\}  -\mu\right)  \right)  } \label{2.10}%
\end{equation}
where $\delta B=\left(  B-\left\langle B\right\rangle \right)  $. The set of
phase variables $\left\{  x\right\}  =\left\{  x_{1},\ldots,x_{N_{e}}\right\}
$ denote the positions and velocities of each electron, e.g., $x_{1}%
\Longleftrightarrow\mathbf{r}_{1},\mathbf{v}_{1}$. The passive dependence on
the ion coordinates has been made explicit here, but will be suppressed in the
following for simplicity of notation, except where needed.

The time dependence of $A(t,\left\{  x\right\}  )$ is generated by the
Hamiltonian (\ref{2.1}) from the initial value $A(\left\{  x\right\}  )$. The
phase functions $A(\left\{  x\right\}  )$ and $B(\left\{  x\right\}  )$ denote
some observables of interest, composed of sums of single particle functions
\begin{equation}
A=\sum_{\alpha=1}^{N_{e}}a(x_{\alpha}),\hspace{0.25in}B=\sum_{\alpha=1}%
^{N_{e}}b(x_{\alpha}). \label{2.11}%
\end{equation}
The special form (\ref{2.11}) allows reduction of the $N_{e}$ electron average
to a corresponding average in the single electron subspace, by partial
integration over $N_{e}-1$ degrees of freedom (see Appendix \ref{appA})%
\begin{equation}
\left\langle A(t)\delta B\right\rangle =\int dxn(\mathbf{r})\phi\left(
v\right)  a(x)\overline{b}(x,t). \label{2.12}%
\end{equation}
Here, $n(\mathbf{r})$ is the equilibrium number density for electrons at a
position $\mathbf{r}$, and $\phi\left(  v\right)  $ is the Maxwell-Boltzmann
velocity distribution. The function $\overline{b}(x,t)$ at $t=0$ is linearly
related to the single particle phase function $b(x)$ in (\ref{2.11})%
\begin{equation}
\overline{b}(x,0)=\overline{b}(x)=b(x)+\int dx^{\prime}n(\mathbf{r}^{\prime
})\phi\left(  v^{\prime}\right)  \left(  g\left(  \mathbf{r},\mathbf{r}%
^{\prime}\right)  -1\right)  b(x^{\prime}), \label{2.13}%
\end{equation}
and $g(\mathbf{r},\mathbf{r}^{\prime})$ is the pair correlation function for
two electrons at $\mathbf{r}$ and $\mathbf{r}^{\prime}$ in the presence of the
ion configuration $\left\{  \mathbf{R}\right\}  $. Finally, the time
dependence of $\overline{b}(x,t)$ is given by the linear kinetic equation in
the single particle phase space
\begin{equation}
\partial_{t}\overline{b}(x,t)+\int dx^{\prime}L\left(  x,x^{\prime};t\right)
\overline{b}(x^{\prime},t)=0, \label{2.15}%
\end{equation}
also derived in Appendix \ref{appA}. All of the results up to this point are
still exact.

\section{Markovian Kinetic Equation and its Relation to DFT}

\label{sec3}The generator of dynamics, $L\left(  x,x^{\prime};t\right)  $,
defined formally in Appendix \ref{appA}, is an appropriate point for the
introduction of approximations. Typically, matter under extreme conditions
does not admit any small parameter expansions because the treatment must
include possible strong Coulomb coupling. Here, a Markov approximation is
chosen that does not prejudice such conditions. Furthermore there is no
scattering context so that the electrons may be free or bound to the ions. A
Markov kinetic equation has a generator whose form does not change in time.
Hence, a practical expression can be determined from $L\left(  x,x^{\prime
};t=0\right)  $ and assumed to hold as well for all later times. One of the
first developments of this idea for\ classical time correlation functions was
given by Lebowitz, Percus, and Sykes \cite{Lebowitz69}, and for the quantum
case by Boercker and Dufty \cite{Boercker81}. The primary difference here is
the presence of the external forces due to the ions. The analysis follows that
of reference \cite{Wrighton08}, and the details are given in Appendix
\ref{appB} with the resulting kinetic equation%

\begin{equation}
\left(  \partial_{t}+\mathbf{v}\cdot\nabla_{\mathbf{r}}-m^{-1}\nabla
_{\mathbf{r}}\mathcal{V}_{ie}\left(  \mathbf{r}\right)  \cdot\nabla
_{\mathbf{v}}\right)  \overline{b}(x,t)=-\mathbf{v}\cdot\nabla_{\mathbf{r}%
}\beta\int dx^{\prime}\mathcal{V}_{ee}\left(  \mathbf{r},\mathbf{r}^{\prime
}\right)  \phi\left(  v^{\prime}\right)  n\left(  \mathbf{r}^{\prime}\right)
\overline{b}(x^{\prime},t). \label{3.1}%
\end{equation}
To understand this result, note that if $\mathcal{V}_{ie}$ and $\mathcal{V}%
_{ee}$ were their (regularized) Coulomb interactions (\ref{3.1}) would be the
random phase approximation (RPA) in the presence of the external ion
potential. Here, however, those potentials have been renormalized by the
initial equilibrium correlations. The left hand side of (\ref{3.1}) describes
single particle motion in an external renormalized ion - electron potential
$\mathcal{V}_{ie}\left(  r\right)  $%
\begin{equation}
\mathcal{V}_{ie}\left(  \mathbf{r},\left\{  \mathbf{R}\right\}  \right)
\equiv-\beta^{-1}\ln n\left(  \mathbf{r},\left\{  \mathbf{R}\right\}  \right)
. \label{3.2}%
\end{equation}
The right side describes dynamical correlations for this single particle
motion with a renormalized electron - electron potential%
\begin{equation}
\mathcal{V}_{ee}\left(  \mathbf{r},\mathbf{r}^{\prime},\left\{  \mathbf{R}%
\right\}  \right)  =-\beta^{-1}c\left(  \mathbf{r},\mathbf{r}^{\prime
},\left\{  \mathbf{R}\right\}  \right)  . \label{3.3}%
\end{equation}
Here $c\left(  \mathbf{r},\mathbf{r}^{\prime},\left\{  \mathbf{R}\right\}
\right)  $ is the electron direct correlation function determined in terms of
$g\left(  \mathbf{r},\mathbf{r}^{\prime},\left\{  \mathbf{R}\right\}  \right)
$ through the Ornstein-Zernicke equation \cite{hansen}, (\ref{a.11}). For weak
coupling, (\ref{3.2}) and (\ref{3.3}) reduce to their Coulomb forms. The
explicit dependence on $\left\{  \mathbf{R}\right\}  $ has been restored at
this point to emphasize that the electron subsystem is non-uniform due to the
presence of the ions.

These renormalizations are due to static correlations of the equilibrium
ensemble and provide the desired connection to DFT. To see this, note that the
electron system is an inhomogeneous electron gas due to the presence of the
ions. The associated equilibrium free energy $F$ is a functional of the
corresponding inhomogeneous density, and is traditionally separated into a
non-interacting part, $F^{(0)},$ and an interacting part, $F^{(1)}$,%
\begin{equation}
F(\beta\mid n)=F^{(0)}(\beta\mid n)+F^{(1)}(\beta\mid n). \label{3.4}%
\end{equation}
The equilibrium density for evaluation of these functionals is determined from
the ion - electron potential by%
\begin{equation}
\frac{\delta F(\beta\mid n)}{\delta n\left(  \mathbf{r},\left\{
\mathbf{R}\right\}  \right)  }=\mu-V_{ei}(\mathbf{r},\left\{  \mathbf{R}%
\right\}  ), \label{3.5}%
\end{equation}
with $V_{ei}(\mathbf{r},\left\{  \mathbf{R}\right\}  )$ given by (\ref{2.2}).
Equation (\ref{3.5}) can be rearranged as%
\begin{equation}
\frac{\delta F^{(0)}(\beta\mid n)}{\delta n\left(  \mathbf{r},\left\{
\mathbf{R}\right\}  \right)  }=\mu-v_{KS}(\mathbf{r},\left\{  \mathbf{R}%
\right\}  ), \label{3.6}%
\end{equation}
where $v_{KS}(\mathbf{r},\left\{  \mathbf{R}\right\}  )$ is known in DFT as
the Kohn-Sham potential%
\begin{equation}
v_{KS}(\mathbf{r},\left\{  \mathbf{R}\right\}  )\equiv V_{ei}\left(
\mathbf{r},\left\{  \mathbf{R}\right\}  \right)  +\frac{\delta F^{(1)}%
(\beta\mid n)}{\delta n\left(  \mathbf{r},\left\{  \mathbf{R}\right\}
\right)  }\;. \label{3.7}%
\end{equation}
Furthermore, for the classical system considered here $F^{(0)}(\beta\mid n)$
can be evaluated exactly to give
\begin{equation}
\frac{\delta F^{(0)}(\beta\mid n)}{\delta n\left(  \mathbf{r},\left\{
\mathbf{R}\right\}  \right)  }=\beta^{-1}\ln n\left(  \mathbf{r},\left\{
\mathbf{R}\right\}  \right)  . \label{3.8}%
\end{equation}
Consequently the renormalized ion - electron potential (\ref{3.2}) becomes%
\begin{equation}
\mathcal{V}_{ie}\left(  \mathbf{r},\left\{  \mathbf{R}\right\}  \right)
=-\frac{\delta F^{(0)}(\beta\mid n)}{\delta n\left(  \mathbf{r},\left\{
\mathbf{R}\right\}  \right)  }=v_{KS}(r,\left\{  \mathbf{R}\right\}  )-\mu.
\label{3.9}%
\end{equation}

With this identification it is instructive to write the kinetic equation
(\ref{3.1}) as%
\begin{equation}
\partial_{t}\overline{b}(,t)-\left\{  h_{KS}\left(  x\right)  ,\overline
{b}(x,t)\right\}  =\frac{1}{\phi\left(  v\right)  n\left(  \mathbf{r}\right)
}\int dx^{\prime}\left\{  \mathcal{V}_{ee}\left(  \mathbf{r},\mathbf{r}%
^{\prime}\right) , \phi\left(  v\right)  n\left(  \mathbf{r}\right)
\phi\left(  v^{\prime}\right)  n\left(  \mathbf{r}^{\prime}\right)
\overline{b}(x^{\prime},t)\right\}  . \label{3.10}%
\end{equation}
Here $\left\{  ,\right\}  $ denotes the Poisson bracket, and $h_{KS}$ is the
Kohn - Sham Hamiltonian,%
\begin{equation}
h_{KS}\left(  x\right)  =\frac{1}{2}mv^{2}+v_{KS}(\mathbf{r}). \label{3.11}%
\end{equation}
As noted in the Introduction a common approximation for evaluating Green -
Kubo time correlation expressions for transport coefficients is the
replacement of the actual Hamiltonian with Coulomb interactions by a sum of
single particle Kohn - Sham Hamiltonians \cite{KG}. The resulting kinetic
theory representation is the same as (\ref{3.10}) with zero on the right side.
The semi-classical analysis here provides an important context for that
approximation, justifying the introduction of the Kohn - Sham dynamics and
making the connection to DFT.

In addition, the renormalized electron - electron potential (\ref{3.3}) is
related to the excess free energy functional by
\begin{equation}
\mathcal{V}_{ee}\left(  \mathbf{r},\mathbf{r}^{\prime},\left\{  \mathbf{R}%
\right\}  \right)  =c\left(  \mathbf{r},\mathbf{r}^{\prime},\left\{
\mathbf{R}\right\}  \right)  =-\frac{\delta^{2}\beta F^{(1)}(\beta,\left\{
\mathbf{R}\right\}  \mid n)}{\delta n\left(  \mathbf{r},\left\{
\mathbf{R}\right\}  \right)  \delta n\left(  \mathbf{r}^{\prime},\left\{
\mathbf{R}\right\}  \right)  }=\frac{\delta\beta v_{KS}(\mathbf{r},\left\{
\mathbf{R}\right\}  )}{\delta n\left(  \mathbf{r}^{\prime},\left\{
\mathbf{R}\right\}  \right)  }. \label{3.12}%
\end{equation}
Thus all of the input for the kinetic equation can be obtained from DFT, even
for conditions of interest for strong coupling.

\section{Dynamic Structure, Response, and Conductivity}

\label{sec4}The correlation functions for the dynamic structure factor,
density response function, and electrical conductivity are closely related.
The dynamic structure factor is determined from the Fourier transform of the
density-density time correlation function%
\begin{equation}
C(\mathbf{r,r}^{\prime},t)=\left\langle \widehat{n}(\mathbf{r},t)\delta
\widehat{n}(\mathbf{r}^{\prime})\right\rangle ,\hspace{0.25in}\widehat
{n}(\mathbf{r})=\sum_{\alpha=1}^{N_{e}}\delta(\mathbf{r}_{\alpha}-\mathbf{r}).
\label{4.1}%
\end{equation}
The density response function is proportional to its time derivative
\cite{McLennan89}%
\begin{equation}
\chi(\mathbf{r,r}^{\prime},t)=\beta\partial_{t}C(\mathbf{r,r}^{\prime},t).
\label{4.2}%
\end{equation}
Finally, using the continuity equation%
\begin{equation}
\partial_{t}\widehat{n}(\mathbf{r},t)+\mathbf{\nabla}\cdot\widehat{\mathbf{j}%
}(\mathbf{r},t)=0,\hspace{0.25in}\widehat{\mathbf{j}}(\mathbf{r})=\sum
_{\alpha=1}^{N_{e}}\delta(\mathbf{r}_{\alpha}-\mathbf{r})\mathbf{v}_{\alpha},
\label{4.3}%
\end{equation}
the response function is related to the current-current correlation function%
\begin{equation}
\partial_{t}\chi(\mathbf{r,r}^{\prime},t)=-\beta\left\langle \mathbf{\nabla
\cdot}\widehat{\mathbf{j}}(\mathbf{r},t)\mathbf{\nabla}^{\prime}\mathbf{\cdot
}\widehat{\mathbf{j}}(\mathbf{r}^{\prime})\right\rangle , \label{4.4}%
\end{equation}
which determines the electrical conductivity.

The general solution to the kinetic equation is given in Appendix \ref{appC}.
When applied to $C(\mathbf{r,r}^{\prime},t)$ and $\chi(\mathbf{r,r}^{\prime
},t)$ integral equations for each are obtained,
\begin{equation}
C(\mathbf{r,r}^{\prime},t)=C_{0}(\mathbf{r,r}^{\prime},t)+\int_{0}^{t}%
d\tau\int d\mathbf{r}^{\prime\prime}\chi_{KG}(\mathbf{r,r}^{\prime\prime
},t-\tau)\int d\mathbf{r}^{\prime\prime\prime}\mathcal{V}_{ee}\left(
\mathbf{r}^{\prime\prime},\mathbf{r}^{\prime\prime\prime}\right)
C(\mathbf{r}^{\prime\prime\prime},\mathbf{r}^{\prime},\tau) \label{4.5}%
\end{equation}
and
\begin{equation}
\chi(\mathbf{r,r}^{\prime},t)=\chi_{KG}(\mathbf{r,r}^{\prime},t)+\int_{0}%
^{t}d\tau\int d\mathbf{r}^{\prime\prime}\chi_{KG}(\mathbf{r,r}^{\prime\prime
},t-\tau)\int d\mathbf{r}^{\prime\prime\prime}\mathcal{V}_{ee}\left(
\mathbf{r}^{\prime\prime},\mathbf{r}^{\prime\prime\prime}\right)
\chi(\mathbf{r}^{\prime\prime\prime},\mathbf{r}^{\prime},\tau) \label{4.6}%
\end{equation}
In each of these the non-interacting Kubo-Greenwood response function occurs%
\begin{equation}
\chi_{KG}(\mathbf{r,r}^{\prime},t-\tau)=-\beta\int d\mathbf{v}\phi\left(
v\right)  n\left(  \mathbf{r}\right)  e^{-\mathcal{L}_{KG}\left(
t-\tau\right)  }\mathbf{v}\cdot\mathbf{\nabla}_{\mathbf{r}}\delta\left(
\mathbf{r}-\mathbf{r}^{\prime}\right)  \label{4.7}%
\end{equation}
Here $\mathcal{L}_{KG}$ is the generator for the Kubo-Greenwood dynamics%
\begin{equation}
\mathcal{L}_{KG}\equiv\mathbf{v}\cdot\nabla_{\mathbf{r}}-m^{-1}\nabla
_{\mathbf{r}}\mathcal{V}_{ie}\left(  \mathbf{r}\right)  \cdot\nabla
_{\mathbf{v}}. \label{4.8}%
\end{equation}
The correlation function $C_{0}(\mathbf{r,r}^{\prime},t)$ has the same
dynamics but also the exact initial conditions%
\begin{equation}
C_{0}(\mathbf{r,r}^{\prime},t)=\int d\mathbf{v}\phi\left(  v\right)  n\left(
\mathbf{r}\right)  e^{-\mathcal{L}_{KG}t}\left(  \delta\left(  \mathbf{r}%
-\mathbf{r}^{\prime}\right)  +n(\mathbf{r}^{\prime})\left(  g\left(
\mathbf{r},\mathbf{r}^{\prime}\right)  -1\right)  \right) . \label{4.9}%
\end{equation}

The solutions to (\ref{4.5}) and (\ref{4.6}) are obtained by first taking
their Laplace transforms%
\begin{equation}
\widetilde{C}(\mathbf{r,r}^{\prime},z)=\int_{0}^{\infty}d\tau e^{-zt}%
C(\mathbf{r,r}^{\prime},t),\hspace{0.25in}\widetilde{\chi}(\mathbf{r,r}%
^{\prime},z)=\int_{0}^{\infty}d\tau e^{-zt}\chi(\mathbf{r,r}^{\prime},t).
\label{4.10}%
\end{equation}
Then the solutions are%
\begin{equation}
\widetilde{C}(\mathbf{r,r}^{\prime},z)=\int d\mathbf{r}^{\prime\prime
}\overline{\epsilon}^{-1}(\mathbf{r,r}^{\prime\prime},z)\widetilde{C}%
_{0}(\mathbf{r}^{\prime\prime}\mathbf{,r}^{\prime},z) \label{4.11}%
\end{equation}
and%
\begin{equation}
\widetilde{\chi}(\mathbf{r,r}^{\prime},z)=\int d\mathbf{r}^{\prime\prime
}\overline{\epsilon}^{-1}(\mathbf{r,r}^{\prime\prime},z)\widetilde{\chi}%
_{KG}(\mathbf{r}^{\prime\prime}\mathbf{,r}^{\prime},z), \label{4.12}%
\end{equation}
where $\overline{\epsilon}^{-1}(\mathbf{r,r}^{\prime\prime},z)$ is the inverse
function for%
\begin{equation}
\overline{\epsilon}(\mathbf{r,r}^{\prime},z)=\delta\left(  \mathbf{r-r}%
^{\prime}\right)  -\int d\mathbf{r}^{\prime\prime}\chi_{KG}(\mathbf{r,r}%
^{\prime\prime},z)\mathcal{V}_{ee}\left(  \mathbf{r}^{\prime\prime}%
,\mathbf{r}^{\prime}\right)  . \label{4.13}%
\end{equation}

\subsection{Dielectric function and local field corrections}

The function $\overline{\epsilon}(\mathbf{r,r}^{\prime},z)$ is closely related
to the dielectric function defined by%

\begin{equation}
\int d\mathbf{r}^{\prime\prime}V(\mathbf{r-r}^{\prime\prime})\widetilde{\chi
}(\mathbf{r}^{\prime\prime}\mathbf{,r}^{\prime},z)=\int d\mathbf{r}%
^{\prime\prime}\epsilon^{-1}(\mathbf{r,r}^{\prime\prime},z)\left(
\epsilon(\mathbf{r}^{\prime\prime}\mathbf{,r}^{\prime},z)-\delta\left(
\mathbf{r}^{\prime\prime}\mathbf{-r}^{\prime}\right)  \right)  . \label{4.14}%
\end{equation}
It has the form%
\begin{equation}
\epsilon(\mathbf{r,r}^{\prime},z)=\delta\left(  \mathbf{r-r}^{\prime}\right)
+\int d\mathbf{r}^{\prime\prime}\int d\mathbf{r}^{\prime\prime\prime
}V(\mathbf{r-r}^{\prime\prime})\chi_{KG}(\mathbf{r}^{\prime\prime}%
\mathbf{,r}^{\prime\prime\prime},z)D^{-1}(\mathbf{r}^{\prime\prime\prime
}\mathbf{,r}^{\prime},z) \label{4.15}%
\end{equation}
with%
\begin{equation}
D(\mathbf{r,r}^{\prime},z)=\delta\left(  \mathbf{r-r}^{\prime}\right)  -\int
d\mathbf{r}^{\prime\prime}\Delta(\mathbf{r,r}^{\prime\prime},z)\chi
_{KG}(\mathbf{r}^{\prime\prime}\mathbf{,r}^{\prime},z). \label{4.16}%
\end{equation}
The leading term on the right side of (\ref{4.16}) gives the random phase
approximation. The second term contains the "dynamic local field corrections"
$\Delta(\mathbf{r,r}^{\prime\prime},z)$. In the present case with
$\widetilde{\chi}(\mathbf{r,r}^{\prime},z)$ given by (\ref{4.12}) the local
field corrections $\Delta(\mathbf{r,r}^{\prime},z)$ are found to be
\begin{equation}
\Delta(\mathbf{r,r}^{\prime},z)\rightarrow\mathcal{V}_{ee}\left(
\mathbf{r},\mathbf{r}^{\prime}\right)  -V(\mathbf{r-r}^{\prime}). \label{4.17}%
\end{equation}
These are the static field corrections due to correlations in the definition
of $\mathcal{V}_{ee}\left(  \mathbf{r},\mathbf{r}^{\prime}\right)  $,
(\ref{3.12})
\begin{equation}
\mathcal{V}_{ee}\left(  \mathbf{r},\mathbf{r}^{\prime}\right)  =-\beta
^{-1}c_{ee}\left(  \mathbf{r},\mathbf{r}^{\prime}\right)  =\frac{\delta
^{2}F_{e}^{(1)}(\beta,\left\{  \mathbf{R}\right\}  \mid n)}{\delta n\left(
\mathbf{r},\left\{  \mathbf{R}\right\}  \right)  \delta n\left(
\mathbf{r}^{\prime},\left\{  \mathbf{R}\right\}  \right)  }. \label{4.18}%
\end{equation}
If the Hartree energy is subtracted from $F_{e}^{(1)}$ the remainder is the
exchange-correlation free energy $F_{xc}$ and
\begin{equation}
\mathcal{V}_{ee}\left(  \mathbf{r},\mathbf{r}^{\prime}\right)  -V(\mathbf{r-r}%
^{\prime})=\frac{\delta^{2}F_{xc}(\beta,\left\{  \mathbf{R}\right\}  \mid
n)}{\delta n\left(  \mathbf{r},\left\{  \mathbf{R}\right\}  \right)  \delta
n\left(  \mathbf{r}^{\prime},\left\{  \mathbf{R}\right\}  \right)  }.
\label{4.19}%
\end{equation}
Thus the static local field corrections are the second functional derivative
of $F_{xc}$.

\subsection{Electrical conductivity}

The frequency dependent electron conductivity is given by its classical
Green-Kubo form \cite{McLennan89}%
\begin{equation}
\sigma\left(  \omega\right)  =\operatorname{Re}\int_{0}^{\infty}dte^{i\omega
t}\psi\left(  t\right)  ,\hspace{0.25in}\psi\left(  t\right)  =\frac{\beta
}{3V}\left\langle \left\langle \widehat{\mathbf{J}}(t)\cdot\widehat
{\mathbf{J}}\right\rangle \right\rangle _{i}. \label{4.20}%
\end{equation}
The double brackets $\left\langle \left\langle {}\right\rangle \right\rangle
_{i}$ denote an average over the electron degrees of freedom, followed by an
average over the ion configurations (see below). The total current
$\widehat{\mathbf{J}}$ is the volume integral of the current density
$\widehat{\mathbf{j}}(\mathbf{r})$
\begin{equation}
\widehat{\mathbf{J}}=\widetilde{\mathbf{j}}(\mathbf{k=0}),\hspace
{0.25in}\widetilde{\mathbf{j}}(\mathbf{k})=\int d\mathbf{re}^{i\mathbf{k\cdot
r}}\widehat{\mathbf{j}}(\mathbf{r}). \label{4.21}%
\end{equation}
The Fourier transform of (\ref{4.4}) gives
\begin{equation}
\partial_{t}\left\langle \chi(\mathbf{k,k}^{\prime},t)\right\rangle _{i}=\beta
k_{m}k_{n}^{\prime}\left\langle \left\langle \widetilde{j}_{m}(\mathbf{k}%
,t)\widetilde{j}_{n}(\mathbf{k}^{\prime})\right\rangle \right\rangle _{i}.
\label{4.22}%
\end{equation}
Once the ion configuration average has been performed the system is isotropic
so this becomes%
\begin{equation}
\partial_{t}\left\langle \chi(\mathbf{k,k}^{\prime},t)\right\rangle
_{i}=-\frac{1}{3}\beta k^{2} \left\langle \left\langle \widetilde{\mathbf{j}%
}(\mathbf{k},t)\cdot\widetilde{\mathbf{j}}(-\mathbf{k})\right\rangle
\right\rangle _{i} \delta_{-k\,k^{\prime}}. \label{4.23}%
\end{equation}
Therefore the current autocorrelation function in the expression for the
conductivity, (\ref{4.20}), is
\begin{equation}
\psi\left(  t\right)  =-\frac{1}{V}\lim_{k\rightarrow0}k^{-2}\partial
_{t}\left\langle \chi(\mathbf{k},-\mathbf{k},t)\right\rangle _{i}
 , \label{4.24}%
\end{equation}
where the response function is given by (\ref{4.12}). If the electron
screening of $\overline{\epsilon}(\mathbf{r,r}^{\prime},z)$ could be neglected
the Kubo-Greenwood model would be obtained,%
\begin{equation}
\psi\left(  t\right)  \rightarrow-\frac{1}{V}\lim_{k\rightarrow0}%
k^{-2}\partial_{t}\left\langle \chi_{KG}(\mathbf{k},-\mathbf{k},t)\right\rangle
_{i}  . \label{4.25}%
\end{equation}

In practice, the conductivity is calculated directly from $\left\langle
\widetilde{\mathbf{j}}(\mathbf{0},t)\cdot\widetilde{\mathbf{j}}(\mathbf{0}%
)\right\rangle $ in the Kubo-Greenwood approximation for each ion
configuration, without reference to the density response function. These
conductivities for the disordered systems are then averaged over all configurations.

\section{Quantum Kinetic Equation}

\label{sec5}In the quantum case the electron correlation function (\ref{2.12})
becomes%
\begin{equation}
\left\langle A(t)\delta B\right\rangle =Tr_{1}f\left(  1\right)
a(1)\overline{b}(1,t), \label{5.1}%
\end{equation}
where the trace is taken over a single electron Hilbert space, and the
classical equilibrium one electron distribution function $n(\mathbf{r}%
)\phi\left(  v\right)  $ has been replaced by its corresponding quantum
operator $f\left(  1\right)  $. Similarly, $a(1)$ and $\overline{b}(1,t)$ are
the operators generalizing the phase space functions $a(x)$ and $\overline
{b}(x,t)$. The analysis of Appendix \ref{appA} follows in an analogous way
\cite{Boercker81}. However, the simplifications of the higher order
correlations from the equilibrium hierarchy, (\ref{b.8}) and (\ref{b.9}), are
more complex. Furthermore, recognition of $\ln n(\mathbf{r})$ as the
functional derivative of the non-interacting free energy functional no longer
applies in the quantum case. Hence, to date the exact evaluation of the
generator for the dynamics $L\left(  t=0\right)  $ in the quantum case has not
been accomplished.

In the meantime an alternative route is to quantize the classical result
derived here. The most direct path is to write the kinetic equation in the
equivalent form%
\begin{equation}
\partial_{t}\overline{b}(1,t)-\left\{  h_{KS}\left(  1\right)  ,\overline
{b}(1,t)\right\}  =f^{-1}\left(  1\right)  Tr_{2}\left\{  \mathcal{V}%
_{ee}\left(  1,2\right)  ,f\left(  1\right)  f\left(  2\right)  \overline
{b}(2,t)\right\}  , \label{5.2}%
\end{equation}
and to quantize it by replacing Poisson brackets by their corresponding
commutators. This gives the operator equation
\begin{equation}
\partial_{t}\overline{b}(1,t)+i\left[  h_{KS}\left(  1\right)  ,\overline
{b}(1,t)\right]  =-f^{-1}\left(  1\right)  Tr_{2}i\left[  \mathcal{V}%
_{ee}\left(  1,2\right)  ,f\left(  1\right)  f\left(  2\right)  \overline
{b}(2,t)\right]  . \label{5.3}%
\end{equation}
The Kohn-Sham Hamiltonian is the operator corresponding to (\ref{3.11}) with
the Kohn-Sham potential determined in the same way as (\ref{3.7}) from the
free energy functional for the quantum system. Similarly, $\mathcal{V}%
_{ee}\left(  1,2\right)  $ is determined from that functional as given in
(\ref{3.3}) and (\ref{3.12}).

Equation (\ref{5.3}) is the quantum random phase approximation for a system of
electrons among a configuration of the ions, with renormalized potentials. Its
classical limit is (\ref{3.1}). The random phase approximation without
renormalization (weak coupling limit) has been established directly for the
quantum case \cite{Boercker81}.

\section{Discussion}

\label{sec6}The objectives here have been two-fold. The first is to describe a
kinetic theory for electrons in a disordered array of ions that is both
practical and free from any assumptions regarding electron-electron or
ion-electron coupling. In particular, the purpose is to do so without the need
for distinction of free and bound electrons. This was accomplished by an exact
evaluation of the generator for time dependence at $t=0$, followed by the
assumption that this generator is a reasonable approximation at all later
times (Markov assumption). The second objective is to make contact between a
controlled many-body theory and the phenomenology of the Kubo-Greenwood model.
This was accomplished by observing that the short time generator has a single
particle dynamics that is the same as the Kubo-Greenwood model, including the
ion-electron force determined from the Kohn-Sham potential of equilibrium DFT.
In addition, the context of that model was exposed, requiring additional
effects of electron-electron screening via a renormalized potential also
determined from DFT. The correlation functions have the structure of the
random phase approximation for an inhomogeneous system, modified by these
potentials from DFT.

Although these conceptual issues of strong coupling, connection to DFT, and
clarification of the Kubo-Greenwood model have been addressed, the practical
application of the kinetic equation developed here is limited by its
semi-classical nature. Some of the most interesting state conditions of warm,
dense matter include strong electron degeneracy, outside the domain of the
regularized quantum potentials assumed here. For these cases the quantum
theory of section \ref{sec5} should be a useful practical tool. The detailed
origin for this equation and its limitations (e.g., absence of
electron-electron collisional effects) will be provided elsewhere.

The semi-classical electrons assumed here nevertheless have an important
domain of validity where degeneracy is weak but electron-electron and
electron-ion coupling can be strong. They have been used in early MD
simulations of hydrogen plasmas \cite{Hansen80}, where electron coupling
strengths of order one were studied at weak to moderate degeneracy. Subsequent
simulations have demonstrated that such potentials can describe the transition
from fully ionized to atomic states, but fail for molecular formation
\cite{Filinov04}. More recently simulations to test the accuracy of different
forms of these effective quantum potentials have been reported
\cite{Whitley15,Benedict12}.

The role of the frozen ion configuration is passive in this analysis 
of the electron dynamics. In practice the systems of interest are ions 
and electrons in which both species are mobile. It is assumed, 
however, that the ions are effectively static on the time scale for 
electron properties. In this case the latter properties are calculated 
as here for a given configuration, and then an average performed over 
configurations. The latter are sampled from an \textit{ab initio} simulation of 
the ions \cite{DFTrev}. A discussion of the effect of ion motion is deferred to 
a later point \cite{Wrighton08}. 

\section{Acknowledgement}

The authors are grateful to S.B. Trickey for his many constructive comments on
an earlier draft. This work was supported by the US DOE Grant DE-SC0002139.

\appendix

\section{Kinetic theory}

\label{appA}

The dynamics of $\left\langle A(t)\delta B\right\rangle _{e}$ is conveniently
expressed in terms of the fundamental correlation function $G(x,x^{\prime};t)$%
\begin{equation}
\left\langle A(t)\delta B\right\rangle =\int dxdx^{\prime}a(x)G(x,x^{\prime
};t)b(x^{\prime}), \label{a.1}%
\end{equation}%
\begin{equation}
G(x,x^{\prime};t)=\left\langle f\left(  x,t\right)  \left(  f\left(
x^{\prime}\right)  -\left\langle f\left(  x^{\prime}\right)  \right\rangle
\right)  \right\rangle ,\hspace{0.3in}f\left(  x\right)  =\sum_{\alpha
=1}^{N_{e}}\delta\left(  x-x_{\alpha}\right)  . \label{a.2}%
\end{equation}
This gives the representation (\ref{2.12})%
\begin{equation}
\left\langle A(t)\delta B\right\rangle =\int dxn(\mathbf{r})\phi\left(
v\right)  a(x)\overline{b}(x,t), \label{a.2a}%
\end{equation}
with the identification%
\begin{equation}
\overline{b}(x,t)=\frac{1}{n(\mathbf{r})\phi(v)}\int dx^{\prime}G(x,x^{\prime
};t)b(x^{\prime}) . \label{a.2b}%
\end{equation}
The initial value is%
\begin{equation}
\overline{b}(x,0)=\overline{b}(x)=\frac{1}{n(\mathbf{r})\phi(v)}\int
dx^{\prime}G(x,x^{\prime};0)b\left(  x^{\prime}\right)  . \label{a.4}%
\end{equation}
Evaluation of $G(x,x^{\prime};0)$ is straightforward from the definitions of
the one and two particle equilibrium distribution functions%
\begin{equation}
n(\mathbf{r}_{1})\phi(v_{1})=\sum_{N_{e}>1}N_{e}Tr_{2,..N}\rho\left(  \left\{
\mathbf{R}\right\}  ,\left\{  x\right\}  \right)  , \label{a.5}%
\end{equation}%
\begin{equation}
n(\mathbf{r}_{1})n(\mathbf{r}_{2})g\left(  \mathbf{r}_{1},\mathbf{r}%
_{2}\right)  \phi(v_{1})\phi(v_{2})=\sum_{N_{e}>2}N_{e}\left(  N_{e}-1\right)
Tr_{3_{e},..N_{e}}\rho\left(  \left\{  \mathbf{R}\right\}  ,\left\{
x\right\}  \right)  , \label{a.6}%
\end{equation}
where $\rho\left(  \left\{  \mathbf{R}\right\}  ,\left\{  x\right\}  \right)
$ is the $N$ electron grand canonical distribution function of (\ref{2.10}).
The result is
\begin{equation}
G(x,x^{\prime};0)=n\left(  \mathbf{r}\right)  \phi\left(  v\right)  \left(
\delta\left(  x-x^{\prime}\right)  +\phi\left(  v^{\prime}\right)  n\left(
\mathbf{r}^{\prime}\right)  h\left(  \mathbf{r},\mathbf{r}^{\prime}\right)
\right)  , \label{a.7}%
\end{equation}
with the hole function defined by $h\left(  \mathbf{r},\mathbf{r}^{\prime
}\right)  =g\left(  \mathbf{r},\mathbf{r}^{\prime}\right)  -1$. This leads to
(\ref{2.13})%
\begin{equation}
\overline{b}(x)=b(x)+\int dx^{\prime}n(\mathbf{r}^{\prime})\phi\left(
v^{\prime}\right)  h\left(  \mathbf{r},\mathbf{r}^{\prime}\right)
b(x^{\prime}). \label{a.8}%
\end{equation}

A formally exact kinetic equation follows from definition of the inverse for
$G(x,x^{\prime};t)$
\begin{equation}
\int dx^{\prime\prime}G^{-1}(x,x^{\prime\prime};t)G(x^{\prime\prime}%
,x^{\prime};t)=\delta\left(  x-x^{\prime}\right)  \label{a.9}%
\end{equation}
and differentiation of (\ref{a.2b})
\begin{align}
\partial_{t}\overline{b}(x,t)  &  =\frac{1}{n(\mathbf{r})\phi(v)}\int
dx^{\prime}\partial_{t}G(x,x^{\prime};t)b(x^{\prime})\nonumber\\
&  =\frac{1}{n(\mathbf{r})\phi(v)}\int dx^{\prime}dx^{\prime\prime}\left(
\partial_{t}G(x,x^{\prime};t)\right)  G^{-1}(x^{\prime},x^{\prime\prime
};t)n(\mathbf{r}^{\prime\prime})\phi(v^{\prime\prime})\overline{b}%
(x^{\prime\prime},t) . \label{a.10}%
\end{align}
The generator for the dynamics is identified as%
\begin{equation}
\partial_{t}\overline{b}(x,t)+\int dx^{\prime}L\left(  x,x^{\prime};t\right)
\overline{b}(x^{\prime},t)=0, \label{a.10a}%
\end{equation}
\begin{equation}
L(x,x^{\prime};t)\equiv-\frac{1}{n(\mathbf{r})\phi(v)}\int dx^{\prime\prime
}\left(  \partial_{t}G(x,x^{\prime\prime};t)\right)  G^{-1}(x^{\prime\prime
},x^{\prime};t)n(\mathbf{r}^{\prime})\phi(v^{\prime}) . \label{a.11}%
\end{equation}

\section{Evaluation of $L(x,x^{\prime};0)$}

\label{appB}The Markov approximation is based on using $L(x,x^{\prime};0)$ as
the generator for dynamics,
\begin{equation}
L(x,x^{\prime};0)=\frac{1}{n(\mathbf{r})\phi(v)}\int dx^{\prime\prime}%
\partial_{t}G(x,x^{\prime\prime};t)\mid_{t=0}G^{-1}(x^{\prime\prime}%
,x^{\prime};0)n(\mathbf{r}^{\prime})\phi(v^{\prime}) . \label{b.1}%
\end{equation}
Consider first $G^{-1}(x^{\prime\prime},x^{\prime};0)$ in the form%
\begin{equation}
G^{-1}(x,x^{\prime};0)=\frac{1}{n\left(  \mathbf{r}\right)  \phi\left(
v\right)  }\delta\left(  x-x^{\prime}\right)  -c\left(  \mathbf{r}%
,\mathbf{r}^{\prime}\right)  . \label{b.2}%
\end{equation}
Then this obeys the inverse condition (\ref{a.9}) with (\ref{a.7}) if
$c\left(  \mathbf{r},\mathbf{r}^{\prime}\right)  $ obeys the equation%
\begin{equation}
c\left(  \mathbf{r},\mathbf{r}^{\prime}\right)  =h\left(  \mathbf{r}%
,\mathbf{r}^{\prime}\right)  -\int d\mathbf{r}^{\prime\prime}h\left(
\mathbf{r},\mathbf{r}^{\prime\prime}\right)  n\left(  \mathbf{r}^{\prime
\prime}\right)  c\left(  \mathbf{r}^{\prime\prime},\mathbf{r}^{\prime}\right)
. \label{b.3}%
\end{equation}
This definition for $c\left(  \mathbf{r},\mathbf{r}^{\prime}\right)  $ is the
Ornstein-Zernicke equation \cite{hansen}.

Next $\partial_{t}G(x,x^{\prime\prime};t)\mid_{t=0}$is evaluated from its
definition (\ref{a.2}) and Newton's equations%
\[
\left(  \partial_{t}+\mathbf{v\cdot\nabla}_{\mathbf{r}}-m^{-1}\nabla
_{\mathbf{r}}V_{ie}\left(  \mathbf{r}\right)  \cdot\nabla_{\mathbf{v}}\right)
f\left(  x,t\right)
\]%
\begin{equation}
=\int dx^{\prime}\left(  \mathbf{\nabla}_{\mathbf{r}}V_{ee}\left(
\mathbf{r},\mathbf{r}^{\prime}\right)  \right)  \cdot m^{-1}\nabla
_{\mathbf{v}}\left(  f\left(  x,t\right)  f\left(  x^{\prime},t\right)
-\delta\left(  x-x^{\prime}\right)  f\left(  x^{\prime},t\right)  \right)  .
\label{b.4}%
\end{equation}
This gives%
\begin{align}
\partial_{t}G(x,x^{\prime};t)  &  \mid_{t=0}=-\left(  \mathbf{v\cdot\nabla
}_{\mathbf{r}}-m^{-1}\nabla_{\mathbf{r}}V_{ie}\left(  \mathbf{r},\right)
\cdot\nabla_{\mathbf{v}}\right)  G(x,x^{\prime};0)\nonumber\\
&  +\int dx^{\prime\prime}\left(  \mathbf{\nabla}_{\mathbf{r}}V_{ee}\left(
\mathbf{r},\mathbf{r}^{\prime\prime}\right)  \right)  \cdot m^{-1}%
\nabla_{\mathbf{v}}\left\langle \left(  f\left(  x\right)  f\left(
x^{\prime\prime}\right)  -\delta\left(  x-x^{\prime\prime}\right)  f\left(
x^{\prime\prime}\right)  \right)  \left(  f\left(  x^{\prime}\right)
-\left\langle f\left(  x^{\prime}\right)  \right\rangle \right)  \right\rangle
. \label{b.4a}%
\end{align}
The average in the second term on the right side can be evaluated using the
expressions%
\begin{equation}
\left\langle f\left(  x\right)  \right\rangle =\phi\left(  v\right)  n\left(
\mathbf{r}\right)  \label{b.4b}%
\end{equation}%
\begin{equation}
\left\langle f\left(  x\right)  f\left(  x^{\prime}\right)  \right\rangle
=\delta\left(  x-x^{\prime}\right)  \phi\left(  v\right)  n\left(
\mathbf{r}\right)  +\phi\left(  v\right)  \phi\left(  v^{\prime}\right)
n\left(  \mathbf{r}\right)  n\left(  \mathbf{r}^{\prime}\right)  g\left(
\mathbf{r},\mathbf{r}^{\prime}\right)  \label{b.5}%
\end{equation}%
\begin{align}
\left\langle \left(  f\left(  x\right)  f\left(  x^{\prime\prime}\right)
-\delta\left(  x-x^{\prime\prime}\right)  f\left(  x^{\prime\prime}\right)
\right)  f\left(  x^{\prime}\right)  \right\rangle  &  =\left(  \delta\left(
x-x^{\prime}\right)  +\delta\left(  x^{\prime\prime}-x^{\prime}\right)
\right)  \phi\left(  v\right)  \phi\left(  v^{\prime\prime}\right)  n\left(
\mathbf{r}\right)  n\left(  \mathbf{r}^{\prime\prime}\right)  g\left(
\mathbf{r},\mathbf{r}^{\prime\prime}\right) \nonumber\\
&  +\phi\left(  v\right)  \phi\left(  v^{\prime}\right)  \phi\left(
v^{\prime\prime}\right)  n\left(  \mathbf{r}\right)  n\left(  \mathbf{r}%
^{\prime}\right)  n\left(  \mathbf{r}^{\prime\prime}\right)  g\left(
\mathbf{r},\mathbf{r}^{\prime},\mathbf{r}^{\prime\prime}\right)  . \label{b.6}%
\end{align}
Here $g\left(  \mathbf{r},\mathbf{r}^{\prime},\mathbf{r}^{\prime\prime
}\right)  $ is defined from the three-electron reduced distribution function
\[
n(\mathbf{r}_{1})n(\mathbf{r}_{2})n(\mathbf{r}_{3})g\left(  \mathbf{r}%
_{1},\mathbf{r}_{2},\mathbf{r}_{3}\right)  \phi(v_{1})\phi(v_{2})\phi(v_{3})
\]%
\begin{equation}
=\sum_{N_{e}>3}N_{e}\left(  N_{e}-1\right)  \left(  N_{e}-2\right)
Tr_{3_{e},..N_{e}}\rho\left(  \left\{  \mathbf{R}\right\}  ,\left\{
x\right\}  \right)  . \label{b.7}%
\end{equation}
Then (\ref{b.4a}) becomes
\begin{align}
\partial_{t}G(x,x^{\prime};t)  &  \mid_{t=0}=-\left(  \mathbf{v\cdot\nabla
}_{\mathbf{r}}-m^{-1}\nabla_{\mathbf{r}}V_{ie}\left(  \mathbf{r}\right)
\cdot\nabla_{\mathbf{v}}\right)  G(x,x^{\prime};0)\nonumber\\
&  -\beta\mathbf{v}\cdot\left(  \mathbf{\nabla}_{\mathbf{r}}V_{ee}\left(
\mathbf{r},\mathbf{r}^{\prime}\right)  \right)  \phi\left(  v\right)
\phi\left(  v^{\prime}\right)  n\left(  \mathbf{r}\right)  n\left(
\mathbf{r}^{\prime}\right)  g\left(  \mathbf{r},\mathbf{r}^{\prime}\right)
\nonumber\\
&  +n\left(  \mathbf{r}\right)  m^{-1}\nabla_{\mathbf{v}}\cdot\delta\left(
x-x^{\prime}\right)  \phi\left(  v\right)  \int d\mathbf{r}^{\prime\prime
}\left(  \mathbf{\nabla}_{\mathbf{r}}V_{ee}\left(  \mathbf{r},\mathbf{r}%
^{\prime\prime}\right)  \right)  n\left(  \mathbf{r}^{\prime\prime}\right)
g\left(  \mathbf{r},\mathbf{r}^{\prime\prime}\right) \nonumber\\
&  -n\left(  \mathbf{r}\right)  n\left(  \mathbf{r}^{\prime}\right)
\phi\left(  v^{\prime}\right)  \phi\left(  v\right)  \beta\mathbf{v}\cdot\int
d\mathbf{r}^{\prime\prime}\left(  \mathbf{\nabla}_{\mathbf{r}}V_{ee}\left(
\mathbf{r},\mathbf{r}^{\prime\prime}\right)  \right)  n\left(  \mathbf{r}%
^{\prime\prime}\right)  \left(  g\left(  \mathbf{r},\mathbf{r}^{\prime
},\mathbf{r}^{\prime\prime}\right)  -g\left(  \mathbf{r},\mathbf{r}%
^{\prime\prime}\right)  \right)  . \label{b.7a}%
\end{align}
The two integrals on the right side can be evaluated using the first two
equations of the BBGKY hierarchy%
\begin{equation}
\int d\mathbf{r}^{\prime\prime}\left(  \mathbf{\nabla}_{\mathbf{r}}%
V_{ee}\left(  \mathbf{r},\mathbf{r}^{\prime\prime}\right)  \right)  n\left(
\mathbf{r}^{\prime\prime}\right)  g\left(  \mathbf{r},\mathbf{r}^{\prime
\prime}\right)  =-\beta^{-1}\mathbf{\nabla}_{\mathbf{r}}\ln n\left(
\mathbf{r}\right)  -\nabla_{\mathbf{r}}V_{ie}\left(  \mathbf{r}\right)  ,
\label{b.8}%
\end{equation}
\[
\int d\mathbf{r}^{\prime\prime}\left(  \mathbf{\nabla}_{\mathbf{r}}%
V_{ee}\left(  \mathbf{r},\mathbf{r}^{\prime\prime}\right)  \right)  n\left(
\mathbf{r}^{\prime\prime}\right)  \left(  g\left(  \mathbf{r},\mathbf{r}%
^{\prime},\mathbf{r}^{\prime\prime}\right)  -g\left(  \mathbf{r}%
,\mathbf{r}^{\prime\prime}\right)  \right)  =-\left(  \mathbf{\nabla
}_{\mathbf{r}}V_{ee}\left(  \mathbf{r}\right)  \right)  g\left(
\mathbf{r},\mathbf{r}^{\prime}\right)
\]%
\begin{equation}
-\left(  \mathbf{\nabla}_{\mathbf{r}}V_{ie}\left(  \mathbf{r}\right)
+\beta^{-1}\mathbf{\nabla}_{\mathbf{r}}\ln n\left(  \mathbf{r}\right)
\right)  h\left(  \mathbf{r},\mathbf{r}^{\prime}\right)  -\beta^{-1}%
\mathbf{\nabla}_{\mathbf{r}}h\left(  \mathbf{r},\mathbf{r}^{\prime}\right)
\label{b.9}%
\end{equation}
to get%
\begin{align}
\partial_{t}G(x,x^{\prime};t)  &  \mid_{t=0}=-\left(  \mathbf{v\cdot\nabla
}_{\mathbf{r}}-m^{-1}\nabla_{\mathbf{r}}V_{ie}\left(  \mathbf{r}\right)
\cdot\nabla_{\mathbf{v}}\right)  G(x,x^{\prime};0)\nonumber\\
&  -n\left(  \mathbf{r}\right)  m^{-1}\nabla_{\mathbf{v}}\cdot\delta\left(
x-x^{\prime}\right)  \phi\left(  v^{\prime}\right)  \left(  \beta
^{-1}\mathbf{\nabla}_{\mathbf{r}}\ln n\left(  \mathbf{r}\right)
+\nabla_{\mathbf{r}}V_{ie}\left(  \mathbf{r}\right)  \right) \nonumber\\
&  +n\left(  \mathbf{r}\right)  n\left(  \mathbf{r}^{\prime}\right)
\phi\left(  v^{\prime}\right)  \phi\left(  v\right)  \beta\mathbf{v}%
\cdot\left(  \mathbf{\nabla}_{\mathbf{r}}V_{ie}\left(  \mathbf{r}%
,\mathbf{r}^{\prime}\right)  +\beta^{-1}\mathbf{\nabla}_{\mathbf{r}}\ln
n\left(  \mathbf{r}\right)  \right)  h\left(  \mathbf{r},\mathbf{r}^{\prime
}\right) \nonumber\\
&  +n\left(  \mathbf{r}\right)  n\left(  \mathbf{r}^{\prime}\right)
\phi\left(  v^{\prime}\right)  \phi\left(  v\right)  \beta\mathbf{v}\cdot
\beta^{-1}\mathbf{\nabla}_{\mathbf{r}}h\left(  \mathbf{r},\mathbf{r}^{\prime
}\right)  . \label{b.10}%
\end{align}
Finally, eliminate the delta function in the second term of the right side
using (\ref{a.5})%

\begin{align}
\partial_{t}G(x,x^{\prime};t)  &  \mid_{t=0}=-\left(  \mathbf{v\cdot\nabla
}_{\mathbf{r}}+m^{-1}\beta^{-1}\mathbf{\nabla}_{\mathbf{r}}\ln n\left(
\mathbf{r}\right)  \cdot\nabla_{\mathbf{v}}\right)  G(x,x^{\prime
};0)\nonumber\\
&  +n\left(  \mathbf{r}\right)  n\left(  \mathbf{r}^{\prime}\right)
\phi\left(  v^{\prime}\right)  \phi\left(  v\right)  \beta\mathbf{v}\cdot
\beta^{-1}\mathbf{\nabla}_{\mathbf{r}}h\left(  \mathbf{r},\mathbf{r}^{\prime
}\right)  . \label{b.12}%
\end{align}
Together with (\ref{b.1}), (\ref{b.2}), and (\ref{b.3}) this gives the desired
result
\begin{align}
L(x,x^{\prime};0)  &  =\left(  \mathbf{v\cdot\nabla}_{\mathbf{r}}-m^{-1}%
\beta^{-1}\mathbf{\nabla}_{\mathbf{r}}\ln n\left(  \mathbf{r},\left\{
\mathbf{R}\right\}  \right)  \cdot\nabla_{\mathbf{v}}\right)  \delta\left(
x-x^{\prime}\right) \nonumber\\
&  -n(\mathbf{r}^{\prime},\left\{  \mathbf{R}\right\}  )\phi(v^{\prime
})\mathbf{v}\cdot\mathbf{\nabla}_{\mathbf{r}}c\left(  \mathbf{r}%
,\mathbf{r}^{\prime},\left\{  \mathbf{R}\right\}  \right)  . \label{b.16}%
\end{align}

\section{Solution to Markov kinetic equation}

\label{appC}A formal solution to the kinetic equation (\ref{3.1}) for
$\overline{b}(x,t)$ is
\begin{equation}
\overline{b}(x,t)=e^{-\mathcal{L}_{KG}t}\overline{b}(x)-\int_{0}^{t}d\tau
e^{-\mathcal{L}_{KG}\left(  t-\tau\right)  }\mathbf{v}\cdot\mathbf{\nabla
}_{\mathbf{r}}\int d\mathbf{r}^{\prime}\beta\mathcal{V}_{ee}\left(
\mathbf{r},\mathbf{r}^{\prime}\right)  I(\mathbf{r}^{\prime},\tau),
\label{c.1}%
\end{equation}
where the generator for the effective single particle (Kubo-Greenwood)
dynamics is%
\begin{equation}
\mathcal{L}_{KG}\equiv\mathbf{v}\cdot\nabla_{\mathbf{r}}-m^{-1}\nabla
_{\mathbf{r}}\mathcal{V}_{ie}\left(  \mathbf{r}\right)  \cdot\nabla
_{\mathbf{v}}, \label{c.2}%
\end{equation}
and the source term $I(\mathbf{r},t)$ is
\begin{equation}
I(\mathbf{r},t)\equiv\int d\mathbf{v}\phi\left(  v\right)  n\left(
\mathbf{r}\right)  \overline{b}(x,t). \label{c.3}%
\end{equation}
Use of (\ref{c.1})--(\ref{c.3}) gives an integral equation for $I(\mathbf{r}%
,t)$,%
\begin{equation}
I(\mathbf{r},t)=I_{0}(\mathbf{r},t)+\int_{0}^{t}d\tau\int d\mathbf{r}^{\prime
}\chi_{KG}(\mathbf{r,r}^{\prime},t-\tau)\int d\mathbf{r}^{\prime\prime
}\mathcal{V}_{ee}\left(  \mathbf{r}^{\prime},\mathbf{r}^{\prime\prime}\right)
I(\mathbf{r}^{\prime\prime},\tau), \label{c.4}%
\end{equation}
with
\begin{equation}
I_{0}(\mathbf{r},t)\equiv\int d\mathbf{v}\phi\left(  v\right)  n\left(
\mathbf{r}\right)  e^{-\mathcal{L}_{KG}t}\overline{b}(x) \label{c.5}%
\end{equation}
and%
\begin{equation}
\chi_{KG}(\mathbf{r,r}^{\prime},t-\tau)=-\beta\int d\mathbf{v}\phi\left(
v\right)  n\left(  \mathbf{r}\right)  e^{-\mathcal{L}_{KG}\left(
t-\tau\right)  }\mathbf{v}\cdot\mathbf{\nabla}_{\mathbf{r}}\delta\left(
\mathbf{r-r}^{\prime}\right)  . \label{c.6}%
\end{equation}

\bigskip

\subsection{Dynamic structure factor}

The dynamic structure factor is determined from the density-density time
correlation function%
\begin{equation}
C(\mathbf{r,r}^{\prime},t)=\left\langle \widehat{n}(\mathbf{r},t)\delta
\widehat{n}(\mathbf{r}^{\prime})\right\rangle , \label{c.7}%
\end{equation}
which corresponds to $a\left(  x_{1}\right)  =\delta\left(  \mathbf{r}%
_{1}\mathbf{-r}\right)  $ and $b\left(  x_{1}\right)  =\delta\left(
\mathbf{r}_{1}\mathbf{-r}^{\prime}\right)  $ in (\ref{2.12}). Then with
(\ref{c.1}) the dynamic structure factor obeys the integral equation%

\begin{equation}
C(\mathbf{r,r}^{\prime},t)=C_{0}(\mathbf{r,r}^{\prime},t)+\int_{0}^{t}%
d\tau\int d\mathbf{r}^{\prime\prime}\chi_{KG}(\mathbf{r,r}^{\prime\prime
},t-\tau)\int d\mathbf{r}^{\prime\prime\prime}\mathcal{V}_{ee}\left(
\mathbf{r}^{\prime\prime},\mathbf{r}^{\prime\prime\prime}\right)
C(\mathbf{r}^{\prime\prime\prime},\mathbf{r}^{\prime},\tau), \label{c.8}%
\end{equation}
where
\begin{equation}
C_{0}(\mathbf{r,r}^{\prime},t)=n(\mathbf{r})\int d\mathbf{v}\phi\left(
v\right)  e^{-\mathcal{L}_{KG}t}\left(  \delta\left(  \mathbf{r-r}^{\prime
}\right)  +n(\mathbf{r}^{\prime})h\left(  \mathbf{r},\mathbf{r}^{\prime
}\right)  \right)  . \label{c.9}%
\end{equation}

The solution to the linear equation (\ref{c.8}) is obtained first by defining
the Laplace transform
\begin{equation}
\widetilde{f}(z)=\int_{0}^{\infty}d\tau e^{-zt}f(t), \label{c.10}%
\end{equation}
to get%

\begin{equation}
\widetilde{C}(\mathbf{r,r}^{\prime},z)=\widetilde{C}_{0}(\mathbf{r,r}^{\prime
},z)+\int d\mathbf{r}^{\prime\prime}\widetilde{\chi}_{KG}(\mathbf{r,r}%
^{\prime\prime},z)\int d\mathbf{r}^{\prime\prime\prime}\mathcal{V}_{ee}\left(
\mathbf{r}^{\prime\prime},\mathbf{r}^{\prime\prime\prime}\right)
\widetilde{C}(\mathbf{r}^{\prime\prime\prime},\mathbf{r}^{\prime},z).
\label{c.11}%
\end{equation}
Next define%

\begin{equation}
\epsilon(\mathbf{r,r}^{\prime},z)=\delta\left(  \mathbf{r-r}^{\prime}\right)
-\int d\mathbf{r}^{\prime\prime}\chi_{KG}(\mathbf{r,r}^{\prime\prime
},z)\mathcal{V}_{ee}\left(  \mathbf{r}^{\prime\prime},\mathbf{r}^{\prime
}\right)  , \label{c.12}%
\end{equation}
so the solution to (\ref{c.11}) is%
\begin{equation}
\widetilde{C}(\mathbf{r,r}^{\prime},z)=\int d\mathbf{r}^{\prime\prime}%
\epsilon^{-1}(\mathbf{r,r}^{\prime\prime},z)\widetilde{C}_{0}(\mathbf{r}%
^{\prime\prime}\mathbf{,r}^{\prime},z) . \label{c.13}%
\end{equation}

\subsection{Density response function}

The density response function is related to the density correlation function
by (\ref{4.2})
\begin{equation}
\chi(\mathbf{r,r}^{\prime},t)=\beta\partial_{t}C(\mathbf{r,r}^{\prime},t).
\label{c.14}%
\end{equation}
Then differentiating (\ref{c.8}) gives%

\begin{align*}
\chi(\mathbf{r,r}^{\prime},t)  &  =\chi_{0}(\mathbf{r,r}^{\prime},t)+\int
d\mathbf{r}^{\prime\prime}\chi_{KG}(\mathbf{r,r}^{\prime\prime},0)\int
d\mathbf{r}^{\prime\prime\prime}\mathcal{V}_{ee}\left(  \mathbf{r}%
^{\prime\prime},\mathbf{r}^{\prime\prime\prime}\right)  C(\mathbf{r}%
^{\prime\prime\prime},\mathbf{r}^{\prime},t)\\
&  -\int_{0}^{t}d\tau\int d\mathbf{r}^{\prime\prime}\beta\partial_{\tau}%
\chi_{KG}(\mathbf{r,r}^{\prime\prime},t-\tau)\int d\mathbf{r}^{\prime
\prime\prime}\mathcal{V}_{ee}\left(  \mathbf{r}^{\prime\prime},\mathbf{r}%
^{\prime\prime\prime}\right)  C(\mathbf{r}^{\prime\prime\prime},\mathbf{r}%
^{\prime},\tau),
\end{align*}%
\begin{align}
&  =\chi_{0}(\mathbf{r,r}^{\prime},t)+\int d\mathbf{r}^{\prime\prime}\chi
_{KG}(\mathbf{r,r}^{\prime\prime},t)\int d\mathbf{r}^{\prime\prime\prime
}\mathcal{V}_{ee}\left(  \mathbf{r}^{\prime\prime},\mathbf{r}^{\prime
\prime\prime}\right)  C(\mathbf{r}^{\prime\prime\prime},\mathbf{r}^{\prime
},0)\nonumber\\
&  +\int_{0}^{t}d\tau\int d\mathbf{r}^{\prime\prime}\chi_{KG}(\mathbf{r,r}%
^{\prime\prime},t-\tau)\int d\mathbf{r}^{\prime\prime\prime}\mathcal{V}%
_{ee}\left(  \mathbf{r}^{\prime\prime},\mathbf{r}^{\prime\prime\prime}\right)
\chi(\mathbf{r}^{\prime\prime\prime},\mathbf{r}^{\prime},\tau), \label{c.14a}%
\end{align}%
\begin{align}
\chi(\mathbf{r,r}^{\prime},t)  &  =\chi_{KG}(\mathbf{r,r}^{\prime
},t)\nonumber\\
&  +\int_{0}^{t}d\tau\int d\mathbf{r}^{\prime\prime}\chi_{KG}(\mathbf{r,r}%
^{\prime\prime},t-\tau)\int d\mathbf{r}^{\prime\prime\prime}\mathcal{V}%
_{ee}\left(  \mathbf{r}^{\prime\prime},\mathbf{r}^{\prime\prime\prime}\right)
\chi(\mathbf{r}^{\prime\prime\prime},\mathbf{r}^{\prime},\tau). \label{c.15}%
\end{align}
The Ornstien-Zernicke equation (\ref{b.3}) has been used to make the identification%

\begin{equation}
\chi_{KG}(\mathbf{r,r}^{\prime},t)=\chi_{0}(\mathbf{r,r}^{\prime},t)+\int
d\mathbf{r}^{\prime\prime}\chi_{KG}(\mathbf{r,r}^{\prime\prime},t)\int
d\mathbf{r}^{\prime\prime\prime}\mathcal{V}_{ee}\left(  \mathbf{r}%
^{\prime\prime},\mathbf{r}^{\prime\prime\prime}\right)  C(\mathbf{r}%
^{\prime\prime\prime},\mathbf{r}^{\prime},0). \label{c.16}%
\end{equation}

Taking the Laplace transform of (\ref{c.15}) gives the solution%
\begin{equation}
\widetilde{\chi}(\mathbf{r,r}^{\prime},z)=\int d\mathbf{r}^{\prime\prime
}\overline{\epsilon}^{-1}(\mathbf{r,r}^{\prime\prime},z)\widetilde{\chi}%
_{KG}(\mathbf{r}^{\prime\prime}\mathbf{,r}^{\prime},z), \label{c.17}%
\end{equation}
where $\overline{\epsilon}^{-1}(\mathbf{r,r}^{\prime},z)$ is the inverse
function associated with%
\begin{equation}
\overline{\epsilon}(\mathbf{r,r}^{\prime},z)=\delta\left(  \mathbf{r-r}%
^{\prime}\right)  -\int d\mathbf{r}^{\prime\prime}\chi_{KG}(\mathbf{r,r}%
^{\prime\prime},z)\mathcal{V}_{ee}\left(  \mathbf{r}^{\prime\prime}%
,\mathbf{r}^{\prime}\right)  . \label{c.18}%
\end{equation}

\bigskip

\bigskip

\bigskip

\end{document}